\documentclass[
reprint,
superscriptaddress,
amsmath,amssymb,
aps,
prl,
]{revtex4-2}

\usepackage{physics}
\usepackage{graphicx}% Include figure files
\usepackage{dcolumn}% Align table columns on decimal point
\usepackage{upgreek}
\usepackage{mathrsfs}
\usepackage{setspace}%Ê¹ÓÃ¼ä¾àºê°ü
\usepackage{bm}
\usepackage[breaklinks=true,colorlinks=true,linkcolor=blue,urlcolor=blue,citecolor=blue]{hyperref}
\allowdisplaybreaks[4]
\usepackage{tikz, mathtools}

\begin{document}

\preprint{APS/123-QED}

\title{Constructing Quantum Many-Body Scars from Hilbert Space Fragmentation}
%\title{Constructing Exact and Approximate Many-Body Scars from Hilbert Space Fragmentation}
%\title{Constructing Quantum Many-Body Scars from Weak Hilbert space Fragmentation}
%\title{Many-Body Scars and Emergent Two-Body Losses from Hilbert Space Fragmentation}

\author{Fan Yang}
\affiliation{Institute for Theoretical Physics, University of Innsbruck, Innsbruck 6020, Austria}
\affiliation{Institute for Quantum Optics and Quantum Information of the Austrian Academy of Sciences, Innsbruck 6020, Austria}

\author{Matteo Magoni}
\affiliation{Institute for Theoretical Physics, University of Innsbruck, Innsbruck 6020, Austria}
\affiliation{Institute for Quantum Optics and Quantum Information of the Austrian Academy of Sciences, Innsbruck 6020, Austria}

\author{Hannes Pichler}
\affiliation{Institute for Theoretical Physics, University of Innsbruck, Innsbruck 6020, Austria}
\affiliation{Institute for Quantum Optics and Quantum Information of the Austrian Academy of Sciences, Innsbruck 6020, Austria}

%\date{\today}

\begin{abstract}
Quantum many-body scars (QMBS) are exotic many-body states that exhibit anomalous non-thermal behavior in an otherwise ergodic system. In this work, we demonstrate a simple, scalable and intuitive construction of QMBS in a kinetically constrained quantum model exhibiting weak Hilbert space fragmentation. We show that exact QMBS can be constructed by injecting a quasiparticle that partially activates the frozen regions in the lattice. Meanwhile, the inelastic collision between multiple quasiparticles allows for the construction of approximate scars, whose damping is governed by an emergent two-body loss. Our findings establish direct connections between quantum many-body scarring and Hilbert space fragmentation, paving the way for systematically constructing exact and approximate QMBS with nontrivial spatial connectivity. The proposed model can be readily implemented in neutral-atom quantum simulators aided by strong Rydberg interactions.
\end{abstract}

\maketitle
	{\it Introduction}.---Generic quantum many-body systems are expected to thermalize at long times according to the eigenstate thermalization hypothesis (ETH) \cite{ETHreview,Borgonovi_2016,Gogolin_2016}. However, there are notable exceptions that violate this paradigm, so that memory of the initial state is also maintained at long times. As a prominent example, quantum many-body scarring provides a mechanism to weakly break ergodicity \cite{Serbyn2021rev}: while most of the eigenstates still satisfy the ETH, non-thermal behavior can be observed by preparing the system in some special initial states \cite{bernien2017probing,turner2018weak,turner2018quantum,Ho2019periodic,Iadecola2019magnon,Khemani2019signatures,lin2019exact,choi2019emergent,Bull2020scars,Zhao2020scars,Turner2021correspondence,Mukherjee2020collapse,Voorden2020scars,bluvstein2021controlling,surace2021exact,PhysRevResearch.6.023146,PhysRevLett.133.190404,lerose2025scars}. Since the discovery of quantum many-body scars (QMBS) in Rydberg quantum simulators \cite{bernien2017probing}, multiple frameworks have been developed to characterize them, such as the spectrum generating algebra \cite{Mark2020unified,Moudgalya2020hubbard,Dea2020tunnels,Pakrouski2020scars,ren2021quasisymmetry}, emergent quasiparticle condensation \cite{Moudgalya2018entanglement,Schecter2019weak}, and projector embeddings \cite{Shiraishi2017systematic,Ok2019topological,kuno2020flat,Omiya2023scars}. Recently, an exotic ergodicity-breaking paradigm---the Hilbert space fragmentation (HSF)---has shown great potential to construct QMBS that are robust to the presence of spatial and temporal randomness \cite{khemani2020localization,moudgalya2022quantum,Chandran2023review}. In a fragmented system, the Hilbert space gets divided into exponentially many dynamically disconnected Krylov subspaces that are not captured by conventional symmetries \cite{pai2019localization,de2019dynamics,sala2020ergodicity,Pai2020fractons,yang2020hilbert,Hudomal2020scars,Moudgalya2021thermalization,zadnik2021foldedI,pozsgay2021integrable,langlett2021hilbert,moudgalya2022hilbert,buca2023unified,yang2025probing}. In the weak version of the HSF \cite{morningstar2020kinetically,francica2023hilbert,wang2023freezing}, a given symmetry sector possesses a primary, ergodic Krylov subspace and a measure-zero set of ETH-violating states. While resembling QMBS, these states are often composed of strictly localized elements. Yet, a comprehensible and scalable construction of nontrivial scar spaces that dynamically connect distant spatial regions remains elusive.

In this work, we discuss a model that highlights the connection between nontrivial QMBS and weak HSF in kinetically constrained spin systems. We identify the key mechanism behind the weak HSF in this model and use it to establish a bottom-up construction of QMBS. Specifically, we find that the measure-zero subspaces and the primary subspace are spanned by states that contain elements with partial and full activity, respectively. By injecting a partially active element into an otherwise frozen state, we can construct exact QMBS that span a polynomially large, spatially extensive scar space, where the dynamics can be mapped to the free propagation of a quasiparticle through the entire lattice. Meanwhile, the inelastic collision between two quasiparticles induces a marginal leakage to the primary subspace, giving rise to approximate scars. We show that the leakage of such approximate QMBS can be described as a two-body loss and efficiently modeled by a Lindblad master equation. The proposed model sheds new light on the construction of nontrivial QMBS from HSF, and can be implemented in existing neutral-atom quantum simulators \cite{Eckner2023realizing,PhysRevLett.130.243001,kim2024realization}.

\begin{figure*}
	\centering
	\includegraphics[width=\linewidth]{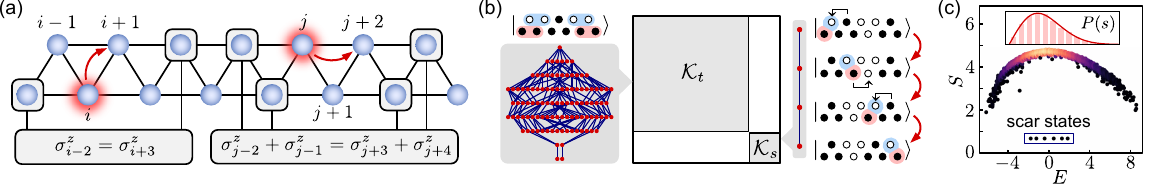}% Here is how to import EPS art
	\caption{(a) Equilateral triangular spin ladder, where the inter-chain and the intra-chain couplings are subjected to the kinetic constraints $\sigma_{i-2}^{z}=\sigma_{i+3}^{z}$ and $\sigma_{j-2}^z+\sigma_{j-1}^z =\sigma_{j+3}^z+\sigma_{j+4}^z$, respectively. (b) Connectivity graphs of the symmetry sector $\{\mathcal{N}_\mathrm{I}=\mathcal{N}_\mathrm{II}=7\}$ for $L=11$, which has two Krylov subspaces $\mathcal{K}_t$ and $\mathcal{K}_s$. (c) Eigenstate entanglement entropy of the symmetry sector $\{\mathcal{N}_\mathrm{I}=\mathcal{N}_\mathrm{II}=10\}$ for $L=17$. The inset shows the level-spacing distribution of the thermal subsector (spatial-inversion symmetric sector), where the solid curve represents the Wigner-Dyson distribution.} \label{fig:fig_model} 
\end{figure*}
	{\it Model}.---We consider a spin-1/2, anisotropic Heisenberg model on an equilateral triangular ladder [see Fig.~\ref{fig:fig_model}(a)]. Keeping only nearest-neighbor (lattice sites separated by one side length) interactions, the Hamiltonian of such a frustrated system reads
	\begin{equation}
		H = \sum_{i}\sum_{r=1,2}\left[J(\sigma_{i}^+ \sigma_{i+r}^- + \sigma_{i}^-\sigma_{i+r}^+) + V n_{i}n_{i+r}\right],\label{eq:H_xxz}
	\end{equation}
	where $J$ and $V$ denote the spin-exchange interaction and the Ising interaction, respectively, with $\sigma_i^\pm=(\sigma_i^x\pm i\sigma_i^y)/2$ and $n_i=(\sigma_i^z + \mathbb{I})/2=|\bullet\rangle\langle\bullet|_i$. Here, $\sigma_i^\alpha$ ($\alpha=x,y,z$) are Pauli operators with respect to the spin-up state $\ket{\bullet}$ and the spin-down state $\ket{\circ}$. This model has a U(1) symmetry corresponding to the conservation of the total number of magnon excitations $\mathcal{N}_\mathrm{I}=\sum_{i} n_{i}$. 
	
	In the highly anisotropic regime $V\gg J$, the total number of magnon bonds (nearest-neighbor magnon excitations) $\mathcal{N}_\mathrm{II}=\sum_{i} (n_in_{i+1}+n_in_{i+2})$ is also conserved, since the creation or annihilation of a bond is far off-resonant and dynamically prohibited. The resulting constrained dynamics can be described by the effective Hamiltonian
	\begin{equation}
		H_\mathrm{eff} = J\sum_{i} \sum_{r=1,2} \mathcal{P}_{i}^{[r]} \left(\sigma_{i}^+ \sigma_{i+r}^- + \sigma_{i}^-\sigma_{i+r}^+\right),\label{eq:H_fxxz}
	\end{equation}
	where the quasilocal projectors $\mathcal{P}_{i}^{[1]}=(\mathbb{I}+\sigma_{i-2}^z\sigma_{i+3}^z)/2$ and $\mathcal{P}_{i}^{[2]}=(5\mathbb{I}+3\sigma_{i-2}^z\sigma_{i-1}^z\sigma_{i+3}^z\sigma_{i+4}^z)/8-[(\sigma_{i-2}^z+\sigma_{i-1}^z)-(\sigma_{i+3}^z+\sigma_{i+4}^z)]^2/16$ impose the kinetic constraints $\sigma_{i-2}^z=\sigma_{i+3}^z$ and $\sigma_{i-2}^z+\sigma_{i-1}^z =\sigma_{i+3}^z+\sigma_{i+4}^z$ on the inter-chain and intra-chain coupling, respectively [see Fig.~\ref{fig:fig_model}(a)]. In the following, we will consider open boundary conditions with $L$ physical sites ($i=1,2,\cdots,L$) \footnote{At both edges, two virtual sites in the state $\ket{\circ}$ are included to account for the boundary effect}.
	
\begin{figure}[b]
		\centering
		\includegraphics[width=\linewidth]{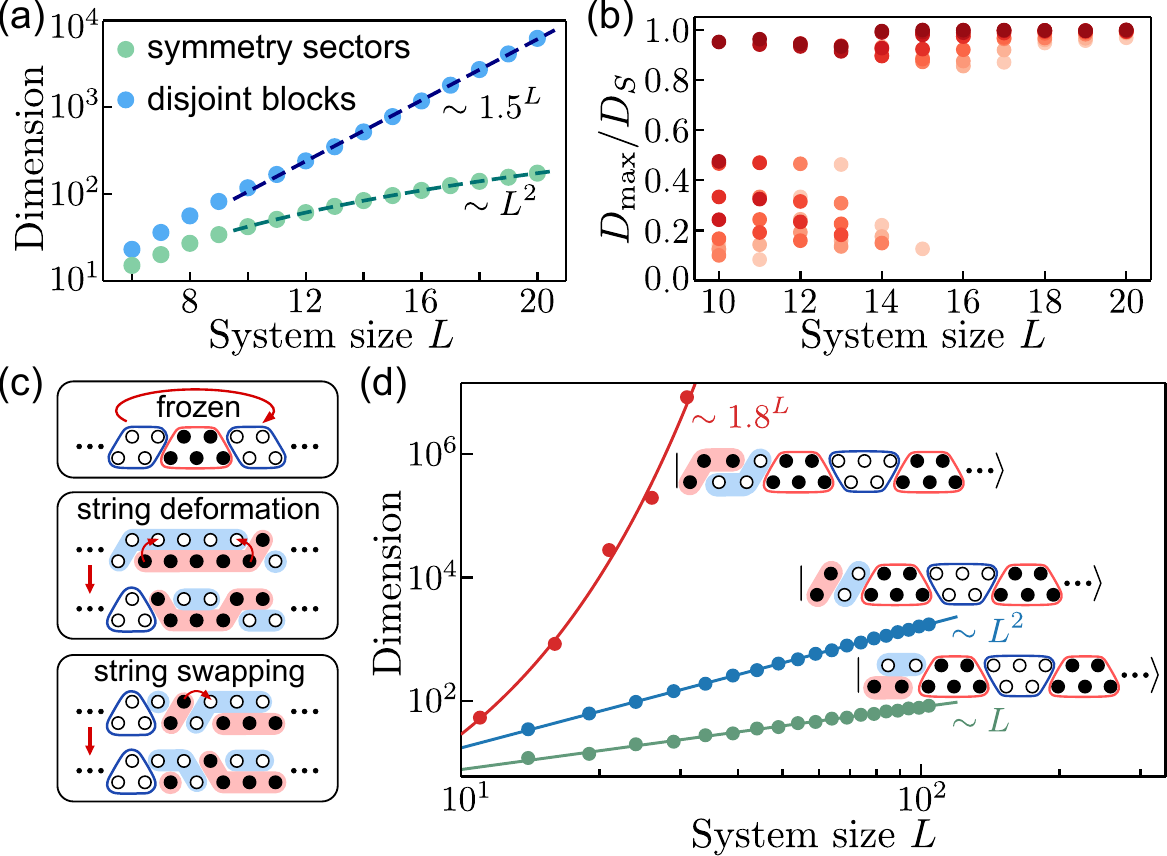}% Here is how to import EPS art
		\caption{(a) Number of distinct symmetry sectors (green dots) and disjoint Krylov subspaces (blue dots) as a function of the system size $L$. (b) Fraction $D_\mathrm{max}/D_S$ for the ten largest fragmented symmetry sectors (ranked and colored according to $D_S$). (c) Elementary patterns (clusters and strings) of the spin configuration and their dynamical behaviors governed by $H_\mathrm{eff}$. (d) Dimension of the Krylov space built from the indicated root state. Scar spaces (blue and green) are constructed by adding a short string to a pattern of frozen clusters.} \label{fig:fig_scaling}
	\end{figure}

	{\it Hilbert space fragmentation}.---We now show that the Hilbert space corresponding to the constrained Hamiltonian~\eqref{eq:H_fxxz} is fragmented. First, we note that the existence of global symmetries implies that the full Hilbert space, spanned by $2^L$ product states $\ket{\psi_i}$ in the computational basis $\{\ket{\bullet}, \ket{\circ}\}$, can be separated into distinct symmetry sectors $\mathcal{S}_{\lambda,\mu}=\mathsf{span}\{\ket{\psi_i}:\mathcal{N}_\mathrm{I}\ket{\psi_i}=\lambda\ket{\psi_i}; \mathcal{N}_\mathrm{II}\ket{\psi_i}=\mu\ket{\psi_i}\}$. In addition, within each symmetry sector, we can dynamically generate disjoint Krylov subspaces $\mathcal{K}_i=\mathsf{span}\{\ket{\psi_i},H_\mathrm{eff}\ket{\psi_i},H_\mathrm{eff}^2\ket{\psi_i},\cdots\}$ by consecutive actions of the Hamiltonian~\eqref{eq:H_fxxz} on each root state $\ket{\psi_i}$. As shown in Fig.~\ref{fig:fig_scaling}(a), the system has a number of Krylov subspaces ($\sim 1.5^L$) which is exponentially larger than the number of symmetry sectors ($\sim L^2$). The lack of obvious conserved quantities characterizing these dynamically disconnected subsectors is then indicative of Hilbert space fragmentation (HSF) \cite{moudgalya2022quantum}.

	To investigate whether the HSF is weak or strong, we calculate the dimension of typical symmetry sectors ($D_S$) and the largest Krylov subspace ($D_\mathrm{max}$) belonging to them. Figure \ref{fig:fig_scaling}(b) shows the ratio $D_\mathrm{max}/D_S$ as a function of the system size for the ten largest fragmented symmetry sectors. For each of these sectors, we can identify a primary Krylov fragment with $D_\mathrm{max}/D_S \rightarrow 1$ as $L$ increases, suggesting that the system is weakly fragmented. The remaining subspaces are of measure-zero, spanned by a frozen state or a small number of states, reminiscent of QMBS.
	
	{\it Construction of exact scars}.---
    We now proceed to develop a systematic construction that allows to understand and characterize these scars in a intuitive fashion. 
    %While each of these subspaces can be regarded as a scar space, the mechanism of the embedding is still not clear, which prevents scalable and systematic constructions of QMBS. To make the dynamics more transparent, we go beyond the constrained-flip-flop description and introduce a more global and intuitive interpretation of the state coupling process. 
    % To systematically construct QMBS from these states in a transparent and scalable way, we adopt an alternative description for the states of the system and its constrained dynamics.
    For this, we first notice that any product-state configuration is composed of two elementary patterns: {\it clusters} and {\it strings}. A {\it cluster} is a domain of $p\geq3$ consecutive sites occupying the same state, e.g., a cluster of magnons ($p=4$) $\ket{\cdots\hspace{2pt}\begin{tikzpicture}[baseline={(0,0)}]\def\size{0.24}\node[draw, circle, inner sep=1.2pt, fill=black] at (0*\size, 0) {};\node[draw, circle, inner sep=1.2pt, fill=black] at (0.5*\size, +0.866*\size) {};    \node[draw, circle, inner sep=1.2pt, fill=black] at (1*\size, 0) {};\node[draw, circle, inner sep=1.2pt, fill=black] at (1.5*\size, +0.866*\size) {};
		\end{tikzpicture}\cdots}$ or a cluster of holes ($p=5$) $\ket{\cdots\hspace{2pt}\begin{tikzpicture}[baseline={(0,0)}]\def\size{0.24}\node[draw, circle, inner sep=1.2pt, fill=white] at (0*\size, 0) {};\node[draw, circle, inner sep=1.2pt, fill=white] at (0.5*\size, +0.866*\size) {};    \node[draw, circle, inner sep=1.2pt, fill=white] at (1*\size, 0) {};\node[draw, circle, inner sep=1.2pt, fill=white] at (1.5*\size, +0.866*\size) {};    \node[draw, circle, inner sep=1.2pt, fill=white] at (2*\size, 0) {};
		\end{tikzpicture}\cdots}$. A {\it string} of length $l$ is a sequence of sites $\{i_1,i_2,\cdots,i_l\}$ occupying the same state, satisfying $0<i_{k+1}-i_k\leq 2$ and $i_{k+2}-i_k\geq 3$, e.g., $\ket{\cdots\hspace{2pt}\begin{tikzpicture}[baseline={(0,0)}]\def\size{0.24}\node[draw, circle, inner sep=1.2pt, fill=black] at (0*\size, 0) {};\node[draw, circle, inner sep=1.2pt, fill=black] at (0.5*\size, +0.866*\size) {};    \node[draw, circle, inner sep=1.2pt, fill=white] at (1*\size, 0) {};\node[draw, circle, inner sep=1.2pt, fill=black] at (1.5*\size, +0.866*\size) {};    \node[draw, circle, inner sep=1.2pt, fill=white] at (2*\size, 0) {};\node[draw, circle, inner sep=1.2pt, fill=black] at (2.5*\size, +0.866*\size) {};    \node[draw, circle, inner sep=1.2pt, fill=black] at (3*\size, 0) {};
		\end{tikzpicture}\cdots}$ contains a magnon-string of length 5 and a hole-string of length 2.

	Clusters and strings have completely different dynamical behavior under the action of $H_\mathrm{eff}$. As depicted in Fig.~\ref{fig:fig_scaling}(c), states composed of only clusters remain frozen, while states containing strings are in general active. Strings support two types of motion: deformation and particle-swapping. First, the inter-chain hoppings $H^{[1]}=J\sum_{i} \mathcal{P}_{i}^{[1]} (\sigma_{i}^+ \sigma_{i+1}^- + \mathrm{H.c.})$ in Eq.~\eqref{eq:H_fxxz} can deform a single string into a different shape, while maintaining the same length \cite{movie}. This term can be rewritten as $H^{[1]}=H_m+H_h$, where $H_m=J\sum_i\ketbra{\circ}_{i-2}(\sigma_i^+\sigma_{i+1}^-+\mathrm{H.c.})\ketbra{\circ}_{i+3}$ and $H_h=J\sum_i\ketbra{\bullet}_{i-2}(\sigma_i^+\sigma_{i+1}^-+\mathrm{H.c.})\ketbra{\bullet}_{i+3}$ induce deformations of magnon-strings and hole-strings, respectively. The active regions of a string include its edges as well as kinks defined as $i_{k+1}-i_k=1$. As we show in the Supplemental Material (SM) \cite{supply}, a single isolated string if length $l$  can deform into exponentially many $\sim[(1+\sqrt{5})/2]^l$ shapes under the action of $H_m$. We also note that deformation of strings of one species (magnons or holes) can effectively break strings of the other species, such that the alternate action of $H_m$ and $H_h$ can generate other configurations where the number of either species of strings is not conserved. In addition to the deformation, strings of the same species can swap elements, e.g., two nearby magnon-strings of length $l$ and $l^\prime\geq 2$ can swap a magnon and evolve into strings of length $l+1$ and $l^\prime-1$. This type of motion is driven by the intra-chain hoppings $H^{[2]}=J\sum_{i} \mathcal{P}_{i}^{[2]} (\sigma_{i}^+ \sigma_{i+2}^- + \mathrm{H.c.})$. We describe in Ref.~\cite{supply} how each constrained flip-flop term of $H^{[2]}$ is related to a swapping process.

	Having interpreted the dynamics from the perspective of string motions, we now study their implications in the fragmentation. For all the fragmented symmetry sectors up to $L=22$, we find that the distribution of string lengths is quite different for the primary Krylov subsectors and the scar spaces: the length $l_\mathrm{max}$ of the longest string in the primary subsectors grows linearly with the system size, but it is bounded ($l_\mathrm{max}\leq3$) for all scar subsectors~\cite{supply}. This may indicate that a long string can significantly activate the system, due to its enhanced complexity and its ability to activate otherwise inert clusters. As support of our conjecture, we consider the scattering between a magnon-string of length $l$ and a magnon-cluster of size $p$. A detailed analysis reveals different rules depending on the length of the string \cite{supply}: (i) A string of $l=1$ (single magnon) attached to a cluster is strictly localized: frozen if $p\geq4$ or locally active if $p=3$. (ii) If an $l\geq 2$ string is attached to a cluster, it can break into a free string of length $(l-1)$ and an isolated cluster of size $(p+1)$. (iii) A free string with $l\leq2$ can fully activate a $p\leq 4$ cluster or partially activate a $p\geq 5$ cluster. (iv) A free string with $l\geq 3$ can fully activate a cluster of arbitrary size. Here, the partial activity refers to the restricted shift of a cluster caused by the string tunneling, in contrast to string-assisted full mobility where the cluster can shift to arbitrary locations \cite{movie}.

	Inspired by the above analysis, we proceed to construct QMBS. The guiding principle is to suppress formation of long strings, as they can fully activate the system. First, rule (i) can be used to construct scars that are locally active \cite{khemani2020localization}, by inserting large, inert clusters between strictly localized string-cluster composites ($l=1$, $p=3$). We are, however, interested in less trivial scars, where distant regions of the system can be connected and entangled. This type of scars can be generated from root states in the ansatz form
	\begin{equation}
		|\psi_\mathrm{root}\rangle=\ket{\boxed{\textrm{short  strings}}\ \boxed{\textrm{scalable\phantom{g}clusters}}}.\label{eq:ansatz}
	\end{equation}
	To ensure that the corresponding symmetry sector is exponentially large, the number of magnon-clusters and hole-clusters needs to be balanced. As a specific choice, we set the scalable clusters to be an ordered state $\ket{\mathbb{Z}_{2p}}$, composed of staggered magnon- and hole-clusters of the same size $p$. In absence of strings, this pattern would be dynamically frozen. The QMBS can then be constructed by adding a short string ($l\leq2$): under the action of Hamiltonian~\eqref{eq:H_fxxz}, the string tunnels through the frozen region ($p\geq5$) as a quasiparticle, giving rise to a polynomially large scar space [rule (iii)]. An exponentially large part of the symmetry sector, i.e., the primary subsector, instead contains long strings that fully activate the clusters [rule (iv)]. Figure \ref{fig:fig_scaling}(d) illustrates examples of Krylov subspaces constructed from $\ket{\mathbb{Z}_{10}}$ states. When an $l=2$ string is attached to the cluster, the generated scar space has dimension $\propto L$. For an $l=2$ free string, the dimension becomes $\propto L^2$. A long string ($l=3$) can no longer generate a scar space, but instead creates a thermal subsector ($\sim 1.8^L$) approaching the size of the corresponding symmetry sector.

	Among all possible constructions, a special one takes place when the root state contains the smallest scalable clusters $|\mathbb{Z}_6\rangle=\ket{\hspace{2pt}\begin{tikzpicture}[baseline={(0,0)}]\def\size{0.24}\node[draw, circle, inner sep=1.2pt, fill=black] at (0*\size, 0) {};\node[draw, circle, inner sep=1.2pt, fill=black] at (0.5*\size, +0.866*\size) {};    \node[draw, circle, inner sep=1.2pt, fill=black] at (1*\size, 0) {};\node[draw, circle, inner sep=1.2pt, fill=white] at (1.5*\size, +0.866*\size) {};    \node[draw, circle, inner sep=1.2pt, fill=white] at (2*\size, 0) {};\node[draw, circle, inner sep=1.2pt, fill=white] at (2.5*\size, +0.866*\size) {};    \node[draw, circle, inner sep=1.2pt, fill=black] at (3*\size, 0) {};\node[draw, circle, inner sep=1.2pt, fill=black] at (3.5*\size, +0.866*\size) {};\node[draw, circle, inner sep=1.2pt, fill=black] at (4*\size, 0) {};\node[draw, circle, inner sep=1.2pt, fill=white] at (4.5*\size, +0.866*\size) {};    \node[draw, circle, inner sep=1.2pt, fill=white] at (5*\size, 0) {};\node[draw, circle, inner sep=1.2pt, fill=white] at (5.5*\size, +0.866*\size) {};
		\end{tikzpicture}\cdots}$. According to rule (iii), any free string can fully activate such set of clusters, so we can only construct QMBS by attaching an $l=1$ string to the first magnon-cluster, which yields the root state
	\begin{equation}
		\ket{\psi_1}=\ket{\hspace{2pt}\begin{tikzpicture}[baseline={(0,0)}]
				\def\size{0.24}
				\draw (-1*\size, 0) -- (-0.5*\size, +0.866*\size);
				\node[draw, circle, inner sep=1.2pt, fill=black] at (-1*\size, 0) {};
				\node[draw, circle, inner sep=1.2pt, fill=white] at (-0.5*\size, +0.866*\size) {};
				\node[draw, circle, inner sep=1.2pt, fill=black] at (0*\size, 0) {};
				\node[draw, circle, inner sep=1.2pt, fill=black] at (0.5*\size, +0.866*\size) {};    \node[draw, circle, inner sep=1.2pt, fill=black] at (1*\size, 0) {};
				\node[draw, circle, inner sep=1.2pt, fill=white] at (1.5*\size, +0.866*\size) {};    \node[draw, circle, inner sep=1.2pt, fill=white] at (2*\size, 0) {};
				\node[draw, circle, inner sep=1.2pt, fill=white] at (2.5*\size, +0.866*\size) {};    \node[draw, circle, inner sep=1.2pt, fill=black] at (3*\size, 0) {};
				\node[draw, circle, inner sep=1.2pt, fill=black] at (3.5*\size, +0.866*\size) {};
				\node[draw, circle, inner sep=1.2pt, fill=black] at (4*\size, 0) {};
				\node[draw, circle, inner sep=1.2pt, fill=white] at (4.5*\size, +0.866*\size) {};    \node[draw, circle, inner sep=1.2pt, fill=white] at (5*\size, 0) {};
				\node[draw, circle, inner sep=1.2pt, fill=white] at (5.5*\size, +0.866*\size) {};
			\end{tikzpicture}\cdots}. \label{eq:qp_states}
	\end{equation}
	For this configuration, while rule (i) sets a localized motion for each individual string-cluster scattering, these processes are connected here, eventually making the magnon-hole pair an effective quasiparticle propagating through the system [see Fig.~\ref{fig:fig_model}(b)]. Formally, these scar states remain frozen under the action of the inter-chain coupling $H^{[1]}$, and are generated by the intra-chain coupling $H^{[2]}$, which can be rewritten as
	$H^{[2]}=J_++J_-$, with $J_+=J\sum_i(\mathcal{P}_{2i}^{[2]}\sigma_{2i}^+\sigma_{2i+2}^-+\mathcal{P}_{2i+1}^{[2]}\sigma_{2i+1}^-\sigma_{2i+3}^+)$ and $J_-=J_+^\dagger$. Application of the raising operator $J_+$ on $|\psi_1\rangle$ then creates a tower of states: 
	\begin{equation}
		\ket{\psi_1}\xrightarrow{J_+}
		\ket{\hspace{2pt}\begin{tikzpicture}[baseline={(0,0)}]
				\def\size{0.24}
				\draw (0.5*\size, +0.866*\size) -- (1*\size, 0);
				\node[draw, circle, inner sep=1.2pt, fill=black] at (-1*\size, 0) {};
				\node[draw, circle, inner sep=1.2pt, fill=black] at (-0.5*\size, +0.866*\size) {};
				\node[draw, circle, inner sep=1.2pt, fill=black] at (0*\size, 0) {};
				\node[draw, circle, inner sep=1.2pt, fill=white] at (0.5*\size, +0.866*\size) {};    \node[draw, circle, inner sep=1.2pt, fill=black] at (1*\size, 0) {};
				\node[draw, circle, inner sep=1.2pt, fill=white] at (1.5*\size, +0.866*\size) {};    \node[draw, circle, inner sep=1.2pt, fill=white] at (2*\size, 0) {};
				\node[draw, circle, inner sep=1.2pt, fill=white] at (2.5*\size, +0.866*\size) {};    \node[draw, circle, inner sep=1.2pt, fill=black] at (3*\size, 0) {};
				\node[draw, circle, inner sep=1.2pt, fill=black] at (3.5*\size, +0.866*\size) {};
				\node[draw, circle, inner sep=1.2pt, fill=black] at (4*\size, 0) {};
				\node[draw, circle, inner sep=1.2pt, fill=white] at (4.5*\size, +0.866*\size) {};    \node[draw, circle, inner sep=1.2pt, fill=white] at (5*\size, 0) {};
				\node[draw, circle, inner sep=1.2pt, fill=white] at (5.5*\size, +0.866*\size) {};
			\end{tikzpicture}\cdots}\xrightarrow{J_+}
		\ket{\hspace{2pt}\begin{tikzpicture}[baseline={(0,0)}]
				\def\size{0.24}
				\draw (2*\size, 0) -- (2.5*\size, +0.866*\size);
				\node[draw, circle, inner sep=1.2pt, fill=black] at (-1*\size, 0) {};
				\node[draw, circle, inner sep=1.2pt, fill=black] at (-0.5*\size, +0.866*\size) {};
				\node[draw, circle, inner sep=1.2pt, fill=black] at (0*\size, 0) {};
				\node[draw, circle, inner sep=1.2pt, fill=white] at (0.5*\size, +0.866*\size) {};    \node[draw, circle, inner sep=1.2pt, fill=white] at (1*\size, 0) {};
				\node[draw, circle, inner sep=1.2pt, fill=white] at (1.5*\size, +0.866*\size) {};    \node[draw, circle, inner sep=1.2pt, fill=black] at (2*\size, 0) {};
				\node[draw, circle, inner sep=1.2pt, fill=white] at (2.5*\size, +0.866*\size) {};    \node[draw, circle, inner sep=1.2pt, fill=black] at (3*\size, 0) {};
				\node[draw, circle, inner sep=1.2pt, fill=black] at (3.5*\size, +0.866*\size) {};
				\node[draw, circle, inner sep=1.2pt, fill=black] at (4*\size, 0) {};
				\node[draw, circle, inner sep=1.2pt, fill=white] at (4.5*\size, +0.866*\size) {};    \node[draw, circle, inner sep=1.2pt, fill=white] at (5*\size, 0) {};
				\node[draw, circle, inner sep=1.2pt, fill=white] at (5.5*\size, +0.866*\size) {};
			\end{tikzpicture}\cdots} \xrightarrow{J_+}\cdots
		.\nonumber
	\end{equation}
	For $L=6k-1$ ($k=2,3,\cdots$), these states form a closed scar space $\mathcal{K}_s=\mathsf{span} \{\ket{\psi_1},J_+\ket{\psi_1},\cdots,(J_+)^M\ket{\psi_1}\}$ with $M=(L-2)/3$. The dynamics can be equivalently described by a tight-binding Hamiltonian $H_S = J\sum_{n=0}^M (c_{n+1}^\dagger c_n+\mathrm{H.c.})$, where we map each magnon-hole-paired state to a hard-core boson: $ (J_+)^n\ket{\psi_1}\leftrightarrow c_n^\dagger \ket{0}$. The corresponding symmetry sector $\{\mathcal{N}_\mathrm{I}= \mathcal{N}_\mathrm{II}=(L+3)/2\}$ is precisely fragmented into this scar space and a thermal subsector $\mathcal{K}_t$, whose connectivity graphs are illustrated in Fig.~\ref{fig:fig_model}(b). We calculate the entanglement entropy $S$ of each eigenstate in the symmetry sector confirming the non-ergodic nature of the scars, despite their location in the middle of the energy spectrum. %We calculate the entanglement entropy $S$ of each eigenstate in the symmetry sector and find a set of high-energy low-entangled eigenstates that are exactly spanned by the scar states, in contrast to the volume-law entangled eigenstates that form a narrow ETH-like band.
    The area-law scaling of the scars also makes them fully distinguishable from the volume-law entangled eigenstates in a narrow ETH-like band [see Fig.~\ref{fig:fig_model}(c)]. Furthermore, the level-spacing statistics follows the Wigner-Dyson distribution [see the inset of Fig.~\ref{fig:fig_model}(c)], ensuring that the thermal subsector is nonintegrable. Altogether, these observations support the construction of a scalable set of exact QMBS.
	
\begin{figure}
		\centering
		\includegraphics[width=\linewidth]{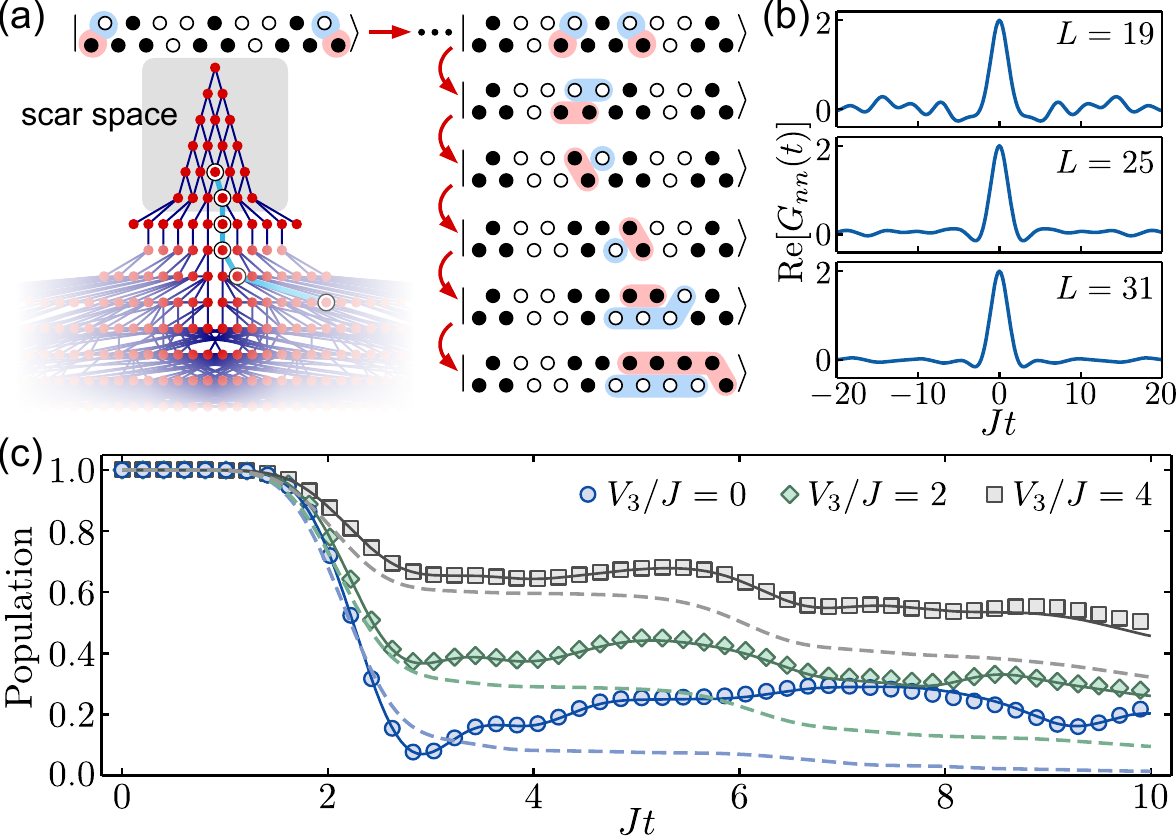}% Here is how to import EPS art
		\caption{(a) Connectivity graph of the symmetry sector $\{\mathcal{N}_\mathrm{I}=\mathcal{N}_\mathrm{II}=11\}$ for $L=19$. The configurations, showing a possible outcome of the inelastic collision, correspond to the vertices that form the highlighted path in the graph. (b) Real part of the correlation function $G_{nn}(t)$ for collisions at the center $n=(L-1)/6$. (c) Evolution of the scar-space population for different interaction strengths $V_3$. The markers are results from exact numerics, the dashed curves show results predicted by the master equation [Eq.~\eqref{eq:master}], and the solid lines are obtained from the non-Markovian model.} \label{fig:fig_con}
	\end{figure}

	{\it Construction of approximate scars}.---We now generalize the previous construction to the situation where multiple quasiparticles are present in the system, which is realized when injecting two magnon-hole pairs at the edges of the $\ket{\mathbb{Z}_6}$ state
	\begin{equation}
		|\psi_2\rangle=\ket{\hspace{2pt}\begin{tikzpicture}[baseline={(0,0)}]
				\def\size{0.24}
				\draw (-1*\size, 0) -- (-0.5*\size, +0.866*\size);
				\node[draw, circle, inner sep=1.2pt, fill=black] at (-1*\size, 0) {};
				\node[draw, circle, inner sep=1.2pt, fill=white] at (-0.5*\size, +0.866*\size) {};
				\node[draw, circle, inner sep=1.2pt, fill=black] at (0*\size, 0) {};
				\node[draw, circle, inner sep=1.2pt, fill=black] at (0.5*\size, +0.866*\size) {};    \node[draw, circle, inner sep=1.2pt, fill=black] at (1*\size, 0) {};
				\node[draw, circle, inner sep=1.2pt, fill=white] at (1.5*\size, +0.866*\size) {};    \node[draw, circle, inner sep=1.2pt, fill=white] at (2*\size, 0) {};
				\node[draw, circle, inner sep=1.2pt, fill=white] at (2.5*\size, +0.866*\size) {};    \node[draw, circle, inner sep=1.2pt, fill=black] at (3*\size, 0) {};
				\node[draw, circle, inner sep=1.2pt, fill=black] at (3.5*\size, +0.866*\size) {};
				\node[draw, circle, inner sep=1.2pt, fill=black] at (4*\size, 0) {};
			\end{tikzpicture}\cdots
			\begin{tikzpicture}[baseline={(0,0)}]
				\def\size{0.24}
				\draw (3.5*\size, +0.866*\size) -- (4*\size, 0);
				\node[draw, circle, inner sep=1.2pt, fill=black] at (-1*\size, 0) {};
				\node[draw, circle, inner sep=1.2pt, fill=black] at (-0.5*\size, +0.866*\size) {};
				\node[draw, circle, inner sep=1.2pt, fill=black] at (0*\size, 0) {};
				\node[draw, circle, inner sep=1.2pt, fill=white] at (0.5*\size, +0.866*\size) {};    \node[draw, circle, inner sep=1.2pt, fill=white] at (1*\size, 0) {};
				\node[draw, circle, inner sep=1.2pt, fill=white] at (1.5*\size, +0.866*\size) {};    \node[draw, circle, inner sep=1.2pt, fill=black] at (2*\size, 0) {};
				\node[draw, circle, inner sep=1.2pt, fill=black] at (2.5*\size, +0.866*\size) {};    \node[draw, circle, inner sep=1.2pt, fill=black] at (3*\size, 0) {};
				\node[draw, circle, inner sep=1.2pt, fill=white] at (3.5*\size, +0.866*\size) {};
				\node[draw, circle, inner sep=1.2pt, fill=black] at (4*\size, 0) {};
			\end{tikzpicture}\hspace{2pt}}.\label{eq:2qp_states}
	\end{equation}
	Application of the constrained Hamiltonian~\eqref{eq:H_fxxz} onto this root state can be interpreted as a quasiparticle collision. To visualize the collision process, we show in Fig.~\ref{fig:fig_con}(a) the connectivity graph generated by $\ket{\psi_2}$. The two quasiparticles move towards each other before colliding. The collision can be either elastic or inelastic. The elastic collision generates a scar space of dimension $\sim L^2$ [shaded tower in Fig.~\ref{fig:fig_con}(a)], spanned by two-quasiparticle states defined through the mapping $\sigma_{3n+1}^x(J_-)^{N-m}\sigma_{3n+1}^x(J_+)^n\ket{\psi_2}\leftrightarrow c_m^\dagger c_n^\dagger \ket{0}$ with $N=(L-1)/3$, where $m,n=0,1,\cdots,N$ ($m> n$) denote the effective site indices of the two hard-core bosons. Interestingly, when the quasiparticles approach each other and become nearest neighbors, a magnon- and hole-string of length 2 are naturally formed. As dictated by rule (iii), these strings can fully activate the frozen state $\ket{\mathbb{Z}_6}$, triggering an inelastic collision where the number of quasiparticles is no longer conserved. Figure \ref{fig:fig_con}(a) illustrates a specific path of such inelastic collision: short strings can rapidly grow into long strings via the inter-chain coupling $H^{[1]}$ and further activate frozen clusters, thereby generating an exponentially large thermal space.

	We give several remarks on this scattering mechanism. First, the quasiparticle mapping here is fundamentally different from those in strongly fragmented models where the scattering is always elastic \cite{de2019dynamics}. This may explain the different physical origins of the strong and the weak fragmentation in these systems. Second, these two-quasiparticle states support approximate scars, as they form an $\mathcal{O}(L^2)$ integrable subspace that has $\mathcal{O}(L)$ local couplings to the thermal space. As shown in the SM \cite{supply}, there is a set of special eigenstates showing atypically large overlap with the scar space, which is a feature of approximate QMBS. In addition, we identify one zero mode induced by many-body caging \cite{tan2025interference,ben2025many,nicolau2025fragmentation,jonay2025localized}. Last, such an inelastic collision provides a way to mimic two-body losses, which can facilitate dissipative preparation of exotic many-body states \cite{rosso2023dynamical,maki2024loss}.
	% $|\mathrm{scar}\rangle=(\sum_{n=0}^{N-2}e^{in\pi}c_n^\dagger c_{n+2}^\dagger |0\rangle+\sigma_1^-\sigma_2^+\ket{\psi_2} +\sigma_{L-1}^-\sigma_{L}^+\ket{\psi_2})/\sqrt{N+2}$
	
	To further elaborate on the last two points, we develop an open-system approach describing the damping of these approximate scars. We first introduce an isomorphism: each state $\ket{\psi}$ generated by $\ket{\psi_2}$ [see Fig.~\ref{fig:fig_con}(a)] is mapped to a state $\ket{\psi_S}\otimes\ket{0}+\ket{0}\otimes\ket{\psi_B}$ that belongs to a tensor-product Hilbert space of quasiparticles ($\ket{\psi_S}$) and an environment ($\ket{\psi_B}$) representing the thermal space. The Hamiltonian~\eqref{eq:H_fxxz} is then reformulated as $H_\mathrm{tot}=H_B+H_S+H_I$, where
	\begin{align}
		H_S=&\sum_n[J(c_{n+1}^\dagger c_{n}+\mathrm{H.c.})+ V_3 c_{n}^\dagger c_{n} - V_3 c_{n+1}^\dagger c_{n}^\dagger c_{n} c_{n+1}], \nonumber\\ 
		H_B=&\sum_k \epsilon_k b_k^\dagger b_k, \quad \textrm{and}\quad H_I=\sum_{k,n} (g_{k,n} b_k^\dagger c_n c_{n+1}  + \mathrm{H.c.})\nonumber 
	\end{align}
	describe the scar-space dynamics, the environment evolution, and the system-environment coupling, respectively. The eigenmode excitation ${b}_k^\dagger\ket{0}$ of the environment corresponds to the $k$-th eigenstate (with eigenenergy $\epsilon_k$) of the Hamiltonian projected onto the thermal space. Here, we include an additional longer-range Ising interaction $V_3\sum_{i}n_{i}n_{i+3}$, which behaves as a nearest-neighbor density interaction between quasiparticles.
	
	In this description, the leakage from the scar space to the environment is mediated by the two-particle loss $\propto c_nc_{n+1}$. By tracing out the environment, we derive the equation of motion for the scar-space density matrix. Introducing the slowly-varying Langevin noise operator $B_n(t) = \sum_{k}g_{k,n}^*b_ke^{-i\delta_kt}$ with $\delta_k=\epsilon_k-2V_3$, the backaction onto the scar space is determined by the reservoir correlation function $G_{mn}(t)=\langle B_m(t) B_n^\dagger(0)\rangle=\sum_k g_{k,m}^*g_{k,n} e^{-i\delta_kt}$. We numerically check that the off-diagonal correlator $G_{mn}(t)|_{m\neq n}\approx 0$ due to the chaotic behavior within the thermal space, while the autocorrelation function $G_{nn}(t)$ converges when increasing the system size due to the locality of the system-bath coupling [see Fig.~\ref{fig:fig_con}(b)]. The rapid decay of $G_{nn}(t)$ suggests that one may apply the Markovian approximation $\mathrm{Re}[G_{nn}(t)]\approx \Gamma_n\delta(t)$, which leads to a Lindblad master equation for the scar-space density matrix $\rho$,
	\begin{align}
		\partial_t\rho=-i[H_S^\prime,\rho]+\sum_n\left(L_n\rho L_n^\dagger- \frac{1}{2}\{L_n^\dagger L_n,\rho\}\right).\label{eq:master}
	\end{align}
	The Lindblad operator $L_n = \sqrt{\Gamma_n}c_nc_{n+1}$ is associated with a two-body loss, and the scar-space Hamiltonian $H_S^\prime=H_S + \sum_n\delta V_nc_{n+1}^\dagger c_n^\dagger c_n c_{n+1}$ is slightly modified by an additional Lamb shift $\delta V_n$. As shown in Fig.~\ref{fig:fig_con}(c), Eq.~\eqref{eq:master} provides a good qualitative description of the inelastic collision. We also note that the leakage from the scar space can be mitigated by increasing the interaction $V_3$, as it induces a repulsion between two quasiparticles. The multi-stage damping of the scar-space population is associated with coherent elastic collisions in the plateau region. To further account for non-Markovian effects, we utilize the pseudo-mode approach \cite{garraway1997nonperturbative,de_vega_review}: each two-body loss $c_n c_{n+1}$ is mediated by the coupling with few auxiliary modes $a_{\sigma,n}$, i.e., $H_n^{(a)}=\sum_{\sigma=1}^{K} (\lambda_{\sigma,n} a_{\sigma,n}^\dagger c_n c_{n+1}  + \mathrm{H.c.})$ \cite{supply}. These embedded modes have distinct detunings $\Delta_{\sigma,n}$ and decay in a Markovian manner via the one-body Lindblad operators $L_{a,n} = \sqrt{\kappa_{\sigma,n}}a_{\sigma,n}$. With optimized parameters, a relatively small set of auxiliary modes ($K=3$) is sufficient to yield quantitative predictions [see the solid lines in Fig.~\ref{fig:fig_con}(c)].

	{\it Conclusion and outlook}.---In conclusion, we presented a kinetically constrained spin model exhibiting weak Hilbert space fragmentation and showed that it originates from the partial activity of mobile elements embedded in a frozen background. This allows us to systematically construct versatile quantum many-body scars, including exact quasiparticle-like scars with different scaling properties, as well as approximate scars whose leakage results from inelastic quasiparticle collisions. Our system can be readily implemented in current quantum simulators based on neutral atoms trapped in an optical lattice \cite{bloch2008review,bakr2009microscope,Landig2016optical,Balewski2014dressing,Zeiher2016interferometry,Jau2016entangling,Gross2017review}. By exciting the atoms to a high-lying Rydberg state, it is possible to engineer interactions that are much stronger than the hopping rate, thereby reaching the desired regime $V \gg J$. Moreover, the recently developed technique of stroboscopic dressing \cite{Hines2023dressing,weckesser2024realization,Cao2025dressing} significantly increases the Rydberg lifetime to hundreds of milliseconds, permitting to experimentally study such systems for tens of tunneling times. As an interesting future direction, one could extend the present framework to models with distinct kinetic constraints \cite{Lan2018slow,Gopalakrishnan2018facilitated,magoni2021bloch,kohlert2023exploring,zhao2024observation,datla2025statistical} or lattice geometries \cite{lee2021frustration,adler2024observation,PhysRevLett.134.010411,buca2025quantum}. Furthermore, the two-body losses identified here can be used as controllable decay channels for engineering driven-dissipative many-body systems \cite{Harrington2022engineered}.

\begin{acknowledgments}
We thank F. Pollmann, M. Knap, S. Moudgalya, A. E. B. Nielsen, and H. Yarloo for helpful discussions. This work is supported by the European Union’s
Horizon Europe research and innovation program under Grant
Agreement No. 101113690 (PASQuanS2.1), the ERC Starting
grant QARA (Grant No. 101041435), and by the Austrian Science Fund
(FWF) (Grant DOI: 10.55776/COE1).
\end{acknowledgments}
	
\bibliography{Reference}
	
\end{document}

\preprint{APS/123-QED}

%\title{Constructing Quantum Many-Body Scars from Weak Hilbert Space Fragmentation}

\title{Supplemental Material for ``Constructing Quantum Many-Body Scars from Hilbert Space Fragmentation''}

\author{Fan Yang}
\affiliation{Institute for Theoretical Physics, University of Innsbruck, Innsbruck 6020, Austria}
\affiliation{Institute for Quantum Optics and Quantum Information of the Austrian Academy of Sciences, Innsbruck 6020, Austria}

\author{Matteo Magoni}
\affiliation{Institute for Theoretical Physics, University of Innsbruck, Innsbruck 6020, Austria}
\affiliation{Institute for Quantum Optics and Quantum Information of the Austrian Academy of Sciences, Innsbruck 6020, Austria}

\author{Hannes Pichler}
\affiliation{Institute for Theoretical Physics, University of Innsbruck, Innsbruck 6020, Austria}
\affiliation{Institute for Quantum Optics and Quantum Information of the Austrian Academy of Sciences, Innsbruck 6020, Austria}

%\date{\today}

\begin{abstract}
In this Supplemental Material, we present several details of the discussion in the main text, including: (i) an intuitive interpretation of the constrained spin dynamics in terms of motions of strings; (ii) the derivation of the total number of configurations taken by a single string; (iii) the scattering between a single string and a frozen cluster; (iv) an open-system model for describing the quasiparticle collision; and (v) approximate scars emerging from the quasiparticle collision.
\end{abstract}

\maketitle

\onecolumngrid
\section{I\lowercase{nterpretation of the dynamics with string motions}}\label{sec:sec1}
In the main text, we mention that all the constrained flip-flop processes can be interpreted as motions of strings. Here, we show explicitly the correspondence. The constrained Hamiltonian $H_\mathrm{eff} = H^{[1]} + H^{[2]}$ contains interchain and intrachain hopping terms
\begin{equation}
	H^{[1]} = J\sum_{i} \mathcal{P}_{i}^{[1]} \left(\sigma_{i}^+ \sigma_{i+1}^- + \sigma_{i}^-\sigma_{i+1}^+\right) \quad \textrm{and}\quad 
	H^{[2]} = J\sum_{i} \mathcal{P}_{i}^{[2]} \left(\sigma_{i}^+ \sigma_{i+2}^- + \sigma_{i}^-\sigma_{i+2}^+\right),
	\label{eq:eq1}
\end{equation}
where $\mathcal{P}_{i}^{[1]}=(\mathbb{I}+\sigma_{i-2}^z\sigma_{i+3}^z)/2$ and $\mathcal{P}_{i}^{[2]}=(5\mathbb{I}+3\sigma_{i-2}^z\sigma_{i-1}^z\sigma_{i+3}^z\sigma_{i+4}^z)/8-[(\sigma_{i-2}^z+\sigma_{i-1}^z)-(\sigma_{i+3}^z+\sigma_{i+4}^z)]^2/16$ impose the kinetic constraints $\sigma_{i-2}^z=\sigma_{i+3}^z$ and $\sigma_{i-2}^z+\sigma_{i-1}^z =\sigma_{i+3}^z+\sigma_{i+4}^z$, respectively. 

\begin{figure}
	\centering
	\includegraphics[width=\linewidth]{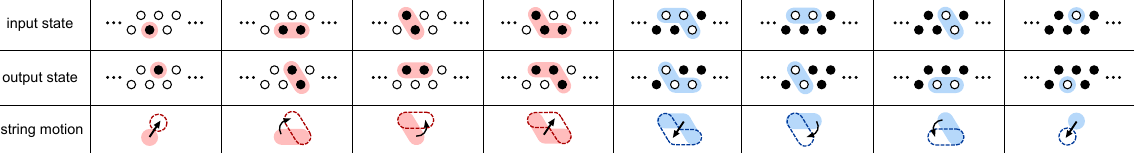}% Here is how to import EPS art
	\caption{Illustration of each term in the interchain coupling $H^{[1]}_i$. The first row depicts the active input state $\ket{\mathrm{input}}$, which evolves into an output state $\ket{\mathrm{output}}=H_i^{[1]}\ket{\mathrm{input}}$ shown in the second row. The third row relates the spin flip-flop process to the deformation of a string, where the dashed curve represents the profile of the output string pattern.} \label{fig:fig_s1} 
\end{figure}

For each individual term $H^{[1]}_i=J\mathcal{P}_{i}^{[1]} \left(\sigma_{i}^+ \sigma_{i+1}^- + \mathrm{H.c.}\right)$ in $H^{[1]}$, $\mathcal{P}_{i}^{[1]}$ projects out $8$ spin configurations at the neighboring sites $\{i-2,i-1,i+2,i+3\}$. The allowed configurations are summarized in Fig.~\ref{fig:fig_s1}, where an input state is transformed into an output state under the action of $H^{[1]}_i$. By identifying the string pattern hidden in each spin configuration, we relate the constrained flip-flop to a specific type of string motion. For example, the first column corresponds to the hopping of an isolated magnon (magnon-string with length 1), and the fourth column describes the deformation of a magnon-string at the kink. To summarize, we find that all the interchain couplings can be interpreted as deformation of a magnon-string governed by $H_m=J\sum_i\ketbra{\circ}_{i-2}(\sigma_i^+\sigma_{i+1}^-+\mathrm{H.c.})\ketbra{\circ}_{i+3}$ and deformation of a hole-string governed by $H_h=J\sum_i\ketbra{\bullet}_{i-2}(\sigma_i^+\sigma_{i+1}^-+\mathrm{H.c.})\ketbra{\bullet}_{i+3}$. We note that the deformation of strings of one species (magnons or holes) can effectively break the other, so that the number of strings of either species is not conserved under the alternate action of $H_m$ and $H_h$.

\begin{figure}
	\centering
	\includegraphics[width=\linewidth]{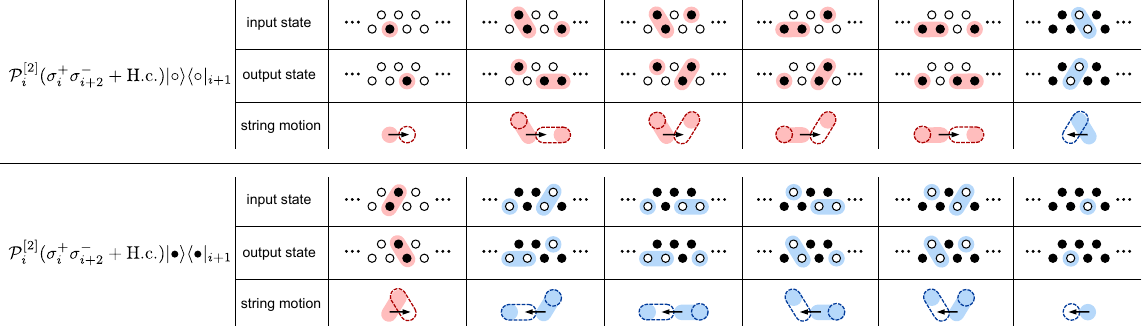}% Here is how to import EPS art
	\caption{Illustration of each individual term in the intrachain constrained Hamiltonian $H^{[2]}_i$ projected onto the state $\ket{\circ}_{i+1}$ and $\ket{\bullet}_{i+1}$, respectively. As in Fig.~\ref{fig:fig_s1}, an active input state is transformed into an output state under the action of the specified Hamiltonian, and each process is related to a pattern of string motions.} \label{fig:fig_s2} 
\end{figure}

For each individual term $H^{[2]}_i=J\mathcal{P}_{i}^{[2]} \left(\sigma_{i}^+ \sigma_{i+2}^- + \mathrm{H.c.}\right)$ in $H^{[2]}$, $\mathcal{P}_{i}^{[2]}$ projects out $10$ spin configurations at the neighboring sites $\{i-2,i-1,i+3,i+4\}$. The remaining $6$ configurations, combined with two possible states at the site $i+1$, yield $12$ terms displayed in Fig.~\ref{fig:fig_s2}. Here, the first and the last column also correspond to the deformation of a single string. However, these couplings do not affect the state connectivity generated by the interchain coupling $H^{[1]}$, as they can be effectively induced by applying $H^{[1]}$ twice. The nontrivial contributions of the intrachain coupling come from the remaining terms, which describe the swapping of particles between two strings: two magnon-strings of length $l$ and $l^\prime\geq 2$ can swap a magnon and evolve into strings of length $l+1$ and $l^\prime-1$ via $H^{[2]}_i\ketbra{\circ}_{i+1}$, while two hole-strings can swap a hole via $H^{[2]}_i\ketbra{\bullet}_{i+1}$.

\section{C\lowercase{ounting the number of configurations taken by a single string}}\label{sec:sec2}
In the main text, we mention that a single isolated magnon-string of length $l$ can be deformed into an arbitrary magnon-string of the same length under the action of $H_m$. In this section, we prove that the total number of these deformations is exponential in $l$, by directly counting the number of string configurations, through a mapping to a simple combinatorial problem.

%Specifically, we compute the total number of ways a string of fixed length $l$ can be inserted in a triangular lattice with $N$ sites, and show that it scales exponentially with $l$. In particular, for $l=N/2 - 1$, this result shows that the number of strings of length $N/2 - 1$ scales exponentially with the system size. This result is relevant since strings of the same length belong to the same Krylov sector, so this number is a lower bound for its dimension. 

%$\ket{\cdots\hspace{2pt}\begin{tikzpicture}[baseline={(0,0)}]\def\size{0.24}\node[draw, circle, inner sep=1.2pt, fill=black] at (0*\size, 0) {};\node[draw, circle, inner sep=1.2pt, fill=black] at (0.5*\size, +0.866*\size) {};    \node[draw, circle, inner sep=1.2pt, fill=white] at (1*\size, 0) {};\node[draw, circle, inner sep=1.2pt, fill=black] at (1.5*\size, +0.866*\size) {};    \node[draw, circle, inner sep=1.2pt, fill=white] at (2*\size, 0) {};\node[draw, circle, inner sep=1.2pt, fill=black] at (2.5*\size, +0.866*\size) {};    \node[draw, circle, inner sep=1.2pt, fill=black] at (3*\size, 0) {}; \end{tikzpicture}\cdots}$

\begin{figure}[b]
	\centering
	\includegraphics[width=0.8\linewidth]{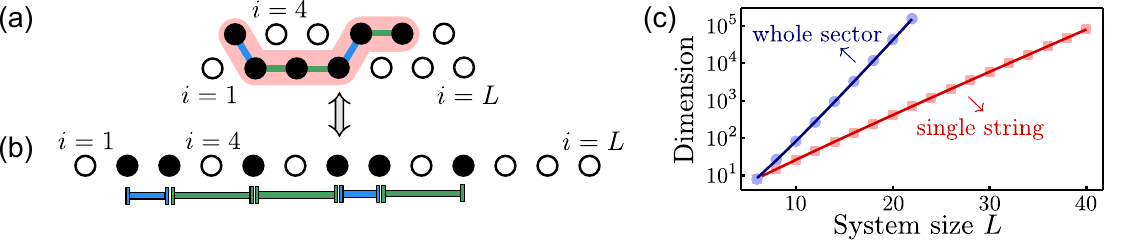}% Here is how to import EPS art
	\caption{(a) and (b) illustrate the mapping between a magnon-string of length $l$ and a sequence of consecutive sticks. Diagonal (horizontal) magnon-bonds in (a) are mapped to short (long) sticks in (b). To correctly represent a string, the sticks must be all attached to each other, with the constraint that no short sticks are nearest neighbors. (c) shows the number of configurations for a single magnon string of length $l=L/2$ (squares) and the dimension of the whole Krylov sector (circles), fitted by the form $0.8\times2^L/L$ and $(2.2+0.32l)\times\phi^{l-1}$, respectively.} \label{fig:fig_s3} 
\end{figure}

A magnon-string of length $l$ has $l$ magnon bonds, which can be either horizontal or diagonal. For example, the string shown in Fig.~\ref{fig:fig_s3}(a) has length $l=5$ and contains three horizontal bonds (in green) and two diagonal bonds (in blue). Moreover, its endpoints are $i=2$ and $i=10$. In order to compute the total number of configurations that such string can reach under the action of $H_m$, it is useful to unfold the triangular ladder into a one-dimensional chain, as shown in Fig.~\ref{fig:fig_s3}(b). In this new frame, bonds are represented by sticks: these can be either one lattice spacing long (diagonal bond) or two lattice spacings long (horizontal bond). Therefore, each magnon-string of length $l$, having $k$ diagonal bonds, $l-k$ horizontal bonds and endpoints $(i_1, i_2)$, corresponds to a sequence of $l$ consecutive sticks, of which $k$ are short and $l-k$ are long, with endpoints $(i_1, i_2)$. It is important to notice that a sequence of sticks must satisfy two conditions in order to represent an isolated magnon-string: \textit{i}) the sticks must be attached to each other and \textit{ii}) short sticks cannot be nearest neighbors. 

Let us now perform the counting. First, for fixed endpoints $(i_1, i_2)$, with $i_2 = i_1 + k + 2(l-k)$, we need to count the number of ways $g(l,k)$ to choose $k$ short sticks among $l$ sticks, with the constraint that no short sticks are nearest neighbors (condition \textit{ii}). By mapping this problem to the one of counting how many binary strings of length $l$ and weight $k$ there are that do not contain neighboring 1s, we get $g(l,k) = \binom{l-k+1}{k}$. In addition, we need to count how many ways we can place a fixed magnon-string, i.e., a fixed sequence of attached (condition \textit{i}) short and long sticks, in the lattice by varying the position of its endpoints. This number is $x(l,k) = L-1 - [k + 2 (l-k)] +1 = L + k -2l$. Therefore, the total number $n(l)$ of configurations taken by a single magnon-string of length $l$ is
\begin{equation}
n(l) = \sum_{k=0}^{\lceil l/2 \rceil} x(l,k) g(l,k) = \sum_{k=0}^{\lceil l/2 \rceil} \left(L + k -2l \right) \binom{l-k+1}{k} \sim p(L,l) \phi^l,
\end{equation}
where $\phi=(1+\sqrt{5})/2$ is the golden ratio and $p(L,l)$ is a polynomial linear in $L$ and $l$. This result shows that the total number of configurations taken by an isolated magnon-string of length $l$ is exponential in $l$. This scaling provides a lower bound to the dimension of the Krylov subspace generated by a single-string root state, since in the counting we are not including the configurations that contain magnon clusters or multiple magnon strings, which can be formed by deformation of the hole strings [see Fig.~\ref{fig:fig_s3}(c)].

\begin{figure}
	\centering
	\includegraphics[width=0.8\linewidth]{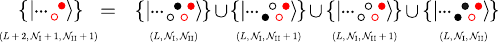}% Here is how to import EPS art
	\caption{Illustration of the first recurrence relation in Eq.~\eqref{eq:recur_rel}. The set of states belonging to the sector $\{\mathcal{N}_\mathrm{I}+1,\mathcal{N}_\mathrm{II}+1\}$ in the lattice with $L+2$ atoms, where the last two are colored in red, is given by the union of four sets of states, each having the two atoms colored in black in one of the four possible configurations. The corresponding quantum numbers defining the sector in the sublattice of the first $L$ sites are indicated below each set.} \label{fig:fig_s4} 
\end{figure}

In fact, we numerically observe that the symmetry sector ($\mathcal{N}_\mathrm{II}=\mathcal{N}_\mathrm{I}-1$) of a single string is fully connected, such that the dimension of the Krylov subspace can be obtained by solving a system of four recurrence relations Eq.~\eqref{eq:recur_rel}. Let us denote with $f(L,\mathcal{N}_\mathrm{I},\mathcal{N}_\mathrm{II})$ the dimension of the symmetry sector $\{\mathcal{N}_\mathrm{I}, \mathcal{N}_\mathrm{II}\}$ in the lattice with $L$ atoms, where $\mathcal{N}_\mathrm{I}$ and $\mathcal{N}_\mathrm{II}$ denote the number of magnon excitations and of magnon bonds, respectively, as defined in the main text. For fixed $L,\mathcal{N}_\mathrm{I}, \mathcal{N}_\mathrm{II}$, the function $f(L,\mathcal{N}_\mathrm{I},\mathcal{N}_\mathrm{II})$ can be decomposed as a sum of four terms, depending on the states of the two sites at the right end of the triangular ladder. In symbols, $f(L,\mathcal{N}_\mathrm{I},\mathcal{N}_\mathrm{II}) = f^{\begin{tikzpicture}[baseline={(0,0)}]\def\size{0.24}\node[draw, circle, inner sep=1.2pt, fill=white] at (0*\size, 0) {};\node[draw, circle, inner sep=1.2pt, fill=black] at (0.3*\size, +0.633*\size) {};
\end{tikzpicture}}(L,\mathcal{N}_\mathrm{I},\mathcal{N}_\mathrm{II}) + f^{\begin{tikzpicture}[baseline={(0,0)}]\def\size{0.24}\node[draw, circle, inner sep=1.2pt, fill=black] at (0*\size, 0) {};\node[draw, circle, inner sep=1.2pt, fill=white] at (0.3*\size, +0.633*\size) {};
\end{tikzpicture}}(L,\mathcal{N}_\mathrm{I},\mathcal{N}_\mathrm{II}) + f^{\begin{tikzpicture}[baseline={(0,0)}]\def\size{0.24}\node[draw, circle, inner sep=1.2pt, fill=white] at (0*\size, 0) {};\node[draw, circle, inner sep=1.2pt, fill=white] at (0.3*\size, +0.633*\size) {};
\end{tikzpicture}}(L,\mathcal{N}_\mathrm{I},\mathcal{N}_\mathrm{II}) + f^{\begin{tikzpicture}[baseline={(0,0)}]\def\size{0.24}\node[draw, circle, inner sep=1.2pt, fill=black] at (0*\size, 0) {};\node[draw, circle, inner sep=1.2pt, fill=black] at (0.3*\size, +0.633*\size) {};
\end{tikzpicture}}(L,\mathcal{N}_\mathrm{I},\mathcal{N}_\mathrm{II})$.
We can write recurrence relations for these functions, since they recursively depend on the same functions with two sites less, as shown in Fig.~\ref{fig:fig_s4}. The recurrence relations read as
\begin{equation}
\begin{cases}
f^{\begin{tikzpicture}[baseline={(0,0)}]\def\size{0.24}\node[draw, circle, inner sep=1.2pt, fill=white] at (0*\size, 0) {};\node[draw, circle, inner sep=1.2pt, fill=black] at (0.3*\size, +0.633*\size) {};
\end{tikzpicture}}(L+2,\mathcal{N}_\mathrm{I}+1,\mathcal{N}_\mathrm{II}+1) &= f^{\begin{tikzpicture}[baseline={(0,0)}]\def\size{0.24}\node[draw, circle, inner sep=1.2pt, fill=white] at (0*\size, 0) {};\node[draw, circle, inner sep=1.2pt, fill=black] at (0.3*\size, +0.633*\size) {};
\end{tikzpicture}}(L,\mathcal{N}_\mathrm{I},\mathcal{N}_\mathrm{II}) + f^{\begin{tikzpicture}[baseline={(0,0)}]\def\size{0.24}\node[draw, circle, inner sep=1.2pt, fill=black] at (0*\size, 0) {};\node[draw, circle, inner sep=1.2pt, fill=white] at (0.3*\size, +0.633*\size) {};
\end{tikzpicture}}(L,\mathcal{N}_\mathrm{I},\mathcal{N}_\mathrm{II}+1) + f^{\begin{tikzpicture}[baseline={(0,0)}]\def\size{0.24}\node[draw, circle, inner sep=1.2pt, fill=white] at (0*\size, 0) {};\node[draw, circle, inner sep=1.2pt, fill=white] at (0.3*\size, +0.633*\size) {};
\end{tikzpicture}}(L,\mathcal{N}_\mathrm{I},\mathcal{N}_\mathrm{II}+1) + f^{\begin{tikzpicture}[baseline={(0,0)}]\def\size{0.24}\node[draw, circle, inner sep=1.2pt, fill=black] at (0*\size, 0) {};\node[draw, circle, inner sep=1.2pt, fill=black] at (0.3*\size, +0.633*\size) {};
\end{tikzpicture}}(L,\mathcal{N}_\mathrm{I},\mathcal{N}_\mathrm{II}) \\
f^{\begin{tikzpicture}[baseline={(0,0)}]\def\size{0.24}\node[draw, circle, inner sep=1.2pt, fill=black] at (0*\size, 0) {};\node[draw, circle, inner sep=1.2pt, fill=white] at (0.3*\size, +0.633*\size) {};
\end{tikzpicture}}(L+2,\mathcal{N}_\mathrm{I}+1,\mathcal{N}_\mathrm{II}+2) &= f^{\begin{tikzpicture}[baseline={(0,0)}]\def\size{0.24}\node[draw, circle, inner sep=1.2pt, fill=white] at (0*\size, 0) {};\node[draw, circle, inner sep=1.2pt, fill=black] at (0.3*\size, +0.633*\size) {};
\end{tikzpicture}}(L,\mathcal{N}_\mathrm{I},\mathcal{N}_\mathrm{II}+1) + f^{\begin{tikzpicture}[baseline={(0,0)}]\def\size{0.24}\node[draw, circle, inner sep=1.2pt, fill=black] at (0*\size, 0) {};\node[draw, circle, inner sep=1.2pt, fill=white] at (0.3*\size, +0.633*\size) {};
\end{tikzpicture}}(L,\mathcal{N}_\mathrm{I},\mathcal{N}_\mathrm{II}+1) + f^{\begin{tikzpicture}[baseline={(0,0)}]\def\size{0.24}\node[draw, circle, inner sep=1.2pt, fill=white] at (0*\size, 0) {};\node[draw, circle, inner sep=1.2pt, fill=white] at (0.3*\size, +0.633*\size) {};
\end{tikzpicture}}(L,\mathcal{N}_\mathrm{I},\mathcal{N}_\mathrm{II}+2) + f^{\begin{tikzpicture}[baseline={(0,0)}]\def\size{0.24}\node[draw, circle, inner sep=1.2pt, fill=black] at (0*\size, 0) {};\node[draw, circle, inner sep=1.2pt, fill=black] at (0.3*\size, +0.633*\size) {};
\end{tikzpicture}}(L,\mathcal{N}_\mathrm{I},\mathcal{N}_\mathrm{II}) \\ 
f^{\begin{tikzpicture}[baseline={(0,0)}]\def\size{0.24}\node[draw, circle, inner sep=1.2pt, fill=white] at (0*\size, 0) {};\node[draw, circle, inner sep=1.2pt, fill=white] at (0.3*\size, +0.633*\size) {};
\end{tikzpicture}}(L+2,\mathcal{N}_\mathrm{I},\mathcal{N}_\mathrm{II}) &= f^{\begin{tikzpicture}[baseline={(0,0)}]\def\size{0.24}\node[draw, circle, inner sep=1.2pt, fill=white] at (0*\size, 0) {};\node[draw, circle, inner sep=1.2pt, fill=black] at (0.3*\size, +0.633*\size) {};
\end{tikzpicture}}(L,\mathcal{N}_\mathrm{I},\mathcal{N}_\mathrm{II}) + f^{\begin{tikzpicture}[baseline={(0,0)}]\def\size{0.24}\node[draw, circle, inner sep=1.2pt, fill=black] at (0*\size, 0) {};\node[draw, circle, inner sep=1.2pt, fill=white] at (0.3*\size, +0.633*\size) {};
\end{tikzpicture}}(L,\mathcal{N}_\mathrm{I},\mathcal{N}_\mathrm{II}) + f^{\begin{tikzpicture}[baseline={(0,0)}]\def\size{0.24}\node[draw, circle, inner sep=1.2pt, fill=white] at (0*\size, 0) {};\node[draw, circle, inner sep=1.2pt, fill=white] at (0.3*\size, +0.633*\size) {};
\end{tikzpicture}}(L,\mathcal{N}_\mathrm{I},\mathcal{N}_\mathrm{II}) + f^{\begin{tikzpicture}[baseline={(0,0)}]\def\size{0.24}\node[draw, circle, inner sep=1.2pt, fill=black] at (0*\size, 0) {};\node[draw, circle, inner sep=1.2pt, fill=black] at (0.3*\size, +0.633*\size) {};
\end{tikzpicture}}(L,\mathcal{N}_\mathrm{I},\mathcal{N}_\mathrm{II}) \\
f^{\begin{tikzpicture}[baseline={(0,0)}]\def\size{0.24}\node[draw, circle, inner sep=1.2pt, fill=black] at (0*\size, 0) {};\node[draw, circle, inner sep=1.2pt, fill=black] at (0.3*\size, +0.633*\size) {};
\end{tikzpicture}}(L+2,\mathcal{N}_\mathrm{I}+2,\mathcal{N}_\mathrm{II}+4) &= f^{\begin{tikzpicture}[baseline={(0,0)}]\def\size{0.24}\node[draw, circle, inner sep=1.2pt, fill=white] at (0*\size, 0) {};\node[draw, circle, inner sep=1.2pt, fill=black] at (0.3*\size, +0.633*\size) {};
\end{tikzpicture}}(L,\mathcal{N}_\mathrm{I},\mathcal{N}_\mathrm{II}+1) + f^{\begin{tikzpicture}[baseline={(0,0)}]\def\size{0.24}\node[draw, circle, inner sep=1.2pt, fill=black] at (0*\size, 0) {};\node[draw, circle, inner sep=1.2pt, fill=white] at (0.3*\size, +0.633*\size) {};
\end{tikzpicture}}(L,\mathcal{N}_\mathrm{I},\mathcal{N}_\mathrm{II}+2) + f^{\begin{tikzpicture}[baseline={(0,0)}]\def\size{0.24}\node[draw, circle, inner sep=1.2pt, fill=white] at (0*\size, 0) {};\node[draw, circle, inner sep=1.2pt, fill=white] at (0.3*\size, +0.633*\size) {};
\end{tikzpicture}}(L,\mathcal{N}_\mathrm{I},\mathcal{N}_\mathrm{II}+3) + f^{\begin{tikzpicture}[baseline={(0,0)}]\def\size{0.24}\node[draw, circle, inner sep=1.2pt, fill=black] at (0*\size, 0) {};\node[draw, circle, inner sep=1.2pt, fill=black] at (0.3*\size, +0.633*\size) {};
\end{tikzpicture}}(L,\mathcal{N}_\mathrm{I},\mathcal{N}_\mathrm{II})
\end{cases}.
\label{eq:recur_rel}
\end{equation}
After deriving the appropriate boundary conditions, these recurrence relations can be numerically solved to provide the dimension $f(L,\mathcal{N}_\mathrm{I},\mathcal{N}_\mathrm{II})$ of the symmetry sector $\{\mathcal{N}_\mathrm{I}, \mathcal{N}_\mathrm{II}\}$ in the lattice with $L$ atoms. 

	\section{S\lowercase{cattering between a string and a cluster}}\label{sec:sec3}
	In the previous section, we show that a string is an active element that can transform into exponentially many configurations. On the other hand, a cluster of magnon excitations remains frozen due to the kinetic constraints. In this section, we investigate the scattering between a string and a cluster, showing how the string can activate the otherwise inert cluster. Specifically, we consider the scattering between a magnon-string of length $l$ and a magnon-cluster of size $p$. The configuration of the system is denoted by $\{l,x_l;p,x_p\}$, where $x_l$ and $x_p$ indicate the location (the starting site index) of the string and the cluster, respectively. A detailed analysis reveals the following rules: 
	
	(i) {\it A string of $l=1$ (a single magnon) directly linked to a cluster has a limited degree of freedom 2 if $p=3$ and becomes frozen if $p\geq4$}. The cases with $p=3$ and $p=4$ are shown below.
	\begin{figure}[!h]
		\centering
		\includegraphics[width=0.6\linewidth]{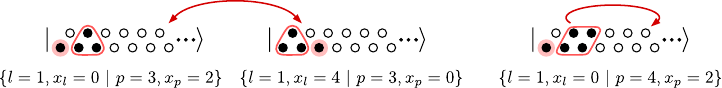}% Here is how to import EPS art
	\end{figure}
	
	(ii) {\it If an $l\geq 2$ string is linked to a cluster, it can break into an active string of length $(l-1)$ and an isolated cluster of size $(p+1)$ via the interchain coupling}. In turn, the state with the free string and the isolated cluster will be subject to additional rules, summarized by rule (iii) and (iv). An example for $\{l=3;p=4\}$ is provided below.
	\begin{figure}[!h]
		\centering
		\includegraphics[width=0.4\linewidth]{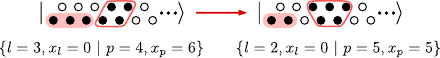}
	\end{figure}

	(iii) {\it A free string with $l\leq2$ can fully activate a $p\leq 4$ cluster and partially activate a $p\geq 5$ cluster}. Here, the full activity means that a string not only can propagate through the cluster (tunneling), making the cluster move $2l$ sites: $\{l,x_l;p,x_p\}\rightarrow \{l,x_l^\prime;p,x_p-2l\}$ ($x_l<x_p$), but also assist the motion of the cluster without tunneling: $\{l,x_l;p,x_p\}\rightarrow \{l,x_l;p,x_p+1\}$. Two examples for $\{l=1;p=4\}$ and $\{l=2;p=4\}$ are shown below.
	\begin{figure}[!h]
		\centering
		\includegraphics[width=\linewidth]{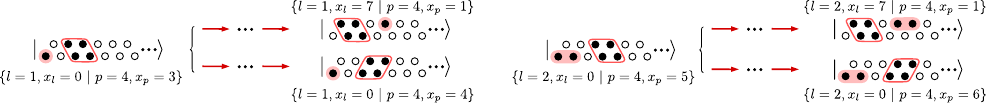}
	\end{figure}
	
	\noindent However, if the cluster is too large ($p\geq 5$), a short string with $l\leq 2$ can only partially activate the cluster through the tunneling mechanism, e.g.,
	\begin{figure}[!h]
		\centering
		\includegraphics[width=\linewidth]{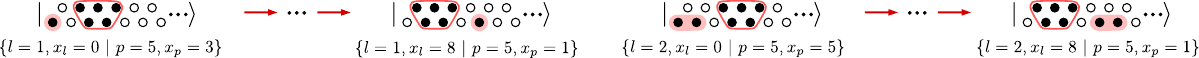}
	\end{figure}
	
	(iv) {\it A free string with $l\geq 3$ can fully activate a cluster of arbitrary size}. Unlike the short string, a long string with $l\geq 3$ can further assist the motion of a cluster of arbitrary size, e.g.,
	\begin{figure}[!h]
		\centering
		\includegraphics[width=0.8\linewidth]{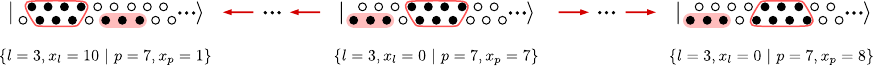} 
	\end{figure}
	
	\noindent The detailed sequence of the steps for the above example is shown in the supplemented movie and can be readily extended to a general initial configuration.
	\begin{figure}[!b]
		\centering
		\includegraphics[width=\linewidth]{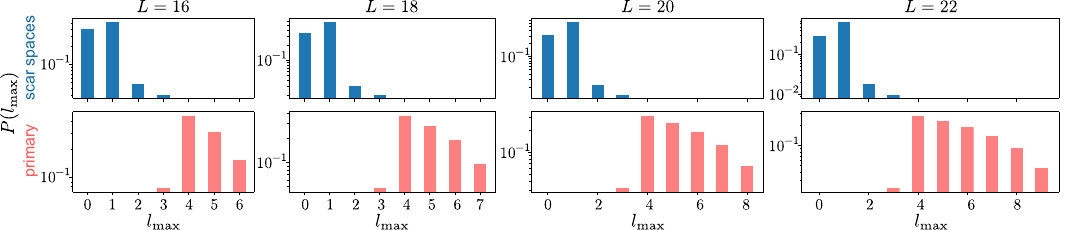}% Here is how to import EPS art
		\caption{Probability distribution $P(l_\mathrm{max})$ of the maximum length $l_\mathrm{max}$ of the string for the scar spaces (blue bars) and the primary subsector (red bars). Here, a Krylov subspace with a dimension $D_\mathrm{K}>0.5D_S$ ($D_S$ being the dimension of the given symmetry sector) is identified as a primary subsector.} \label{fig:fig_s5} 
	\end{figure}	
    
	The above rules allow us to construct scalable scar states. The general idea is to suppress the activity of the system, or equivalently, to suppress the formation of long strings that can fully activate the system. First, we can construct scars by concatenating strictly localized elements indicated in rule (i), e.g., choosing a root state
	\begin{equation}
\ket{\psi_\mathrm{root}}=\ket{\hspace{2pt}\begin{tikzpicture}[baseline={(0,0)}]
				\def\size{0.24}
				\node[draw, circle, inner sep=1.2pt, fill=black] at (-1*\size, 0) {};
				\node[draw, circle, inner sep=1.2pt, fill=white] at (-0.5*\size, +0.866*\size) {};
				\node[draw, circle, inner sep=1.2pt, fill=black] at (0*\size, 0) {};
				\node[draw, circle, inner sep=1.2pt, fill=black] at (0.5*\size, +0.866*\size) {};    \node[draw, circle, inner sep=1.2pt, fill=black] at (1*\size, 0) {};
				\node[draw, circle, inner sep=1.2pt, fill=white] at (1.5*\size, +0.866*\size) {};    \node[draw, circle, inner sep=1.2pt, fill=white] at (2*\size, 0) {};
				\node[draw, circle, inner sep=1.2pt, fill=white] at (2.5*\size, +0.866*\size) {};    \node[draw, circle, inner sep=1.2pt, fill=white] at (3*\size, 0) {};
				\node[draw, circle, inner sep=1.2pt, fill=white] at (3.5*\size, +0.866*\size) {};
				\node[draw, circle, inner sep=1.2pt, fill=black] at (4*\size, 0) {};
				\node[draw, circle, inner sep=1.2pt, fill=white] at (4.5*\size, +0.866*\size) {};    \node[draw, circle, inner sep=1.2pt, fill=black] at (5*\size, 0) {};
				\node[draw, circle, inner sep=1.2pt, fill=black] at (5.5*\size, +0.866*\size) {};
				\node[draw, circle, inner sep=1.2pt, fill=black] at (6*\size, 0) {};
				\node[draw, circle, inner sep=1.2pt, fill=white] at (6.5*\size, +0.866*\size) {};
				\node[draw, circle, inner sep=1.2pt, fill=white] at (7*\size, 0) {};
				\node[draw, circle, inner sep=1.2pt, fill=white] at (7.5*\size, +0.866*\size) {};
				\node[draw, circle, inner sep=1.2pt, fill=white] at (8*\size, 0) {};
				\node[draw, circle, inner sep=1.2pt, fill=white] at (8.5*\size, +0.866*\size) {};
				\node[draw, circle, inner sep=1.2pt, fill=black] at (9*\size, 0) {};
				\node[draw, circle, inner sep=1.2pt, fill=white] at (9.5*\size, +0.866*\size) {};
				\node[draw, circle, inner sep=1.2pt, fill=black] at (10*\size, 0) {};
				\node[draw, circle, inner sep=1.2pt, fill=black] at (10.5*\size, +0.866*\size) {};
				\node[draw, circle, inner sep=1.2pt, fill=black] at (11*\size, 0) {};
			\end{tikzpicture}\cdots}. \label{eq:III1}
	\end{equation}
	The generated scar space only has local mobility, where different regions of the system are uncorrelated. To construct more nontrivial scars where different regions of the system are connected, we can make use of rule (i) and rule (iii) to construct a scar space generated by a root state in the form
	\begin{equation}
		|\psi_\mathrm{root}\rangle=\ket{\boxed{\textrm{short  strings}}\ \boxed{\textrm{scalable\phantom{g}clusters}}}.\label{eq:III2}
	\end{equation}
	Here, the scalable clusters can be chosen in the form of staggered magnon- and hole-clusters as in the main text. In this type of construction, a short string only partially activates the system by propagating through the clusters, exhibiting quasiparticle behaviors. For example, the root state considered in Eq.~(4) of the main text generates scar states that have an intuitive single-quasiparticle mapping. Alternatively, guided by rule (iii), a short string ($l\leq2$) partially activates the system composed of large clusters ($p\geq5$). These types of scars have either single- or two-quasiparticle mapping and are illustrated in Fig.~2(d) of the main text. Here, the string needs to be short because rule (iv) dictates that a long string can fully activate arbitrary clusters. This fact is also illustrated in Fig.~2(d).

	Although the above discussion only involves the scattering between a single string and a cluster, some of the rules might be general. For example, we numerically verify that (up to $L=22$) all the fragmented symmetry sectors are composed of short strings. Figure \ref{fig:fig_s5} shows the distribution of the maximum length of strings $l_\mathrm{max}$ for the primary Krylov subsector and the scar space. As the system size increases, $l_\mathrm{max}$ becomes proportionally large for the primary subsectors, but is bounded ($l_\mathrm{max}\leq3$) for all the scar subsectors. Based on the analysis of the string-cluster scattering as well as the numerical evidence here, we conjecture that a free long string with $l\geq 3$ can always generate a primary Krylov subsector approaching the size of the given symmetry sector up to the thermodynamic limit.

	\section{O\lowercase{pen-system approach for the inelastic quasiparticle collision}}\label{sec:sec4}
	With the mapping introduced in the main text, the inelastic quasiparticle collision is reformulated as an open-system problem, whose Hamiltonian is given by $H_\mathrm{tot}=H_B+H_S+H_I$, where
	\begin{equation}
		H_S=\sum_n[J(c_{n+1}^\dagger c_{n}+\mathrm{H.c.})+ V_3 c_{n}^\dagger c_{n} - V_3 c_{n+1}^\dagger c_{n}^\dagger c_{n} c_{n+1}],\
		H_B=\sum_k \epsilon_k b_k^\dagger b_k, \textrm{ and } H_I=\sum_{k,n} (g_{k,n} b_k^\dagger c_n c_{n+1}  + \mathrm{H.c.}) \nonumber
	\end{equation}
	describe the system (scar-space) dynamics, the environment (thermal-space) evolution, and the system-environment coupling, respectively. In the interaction picture, the evolution of the full density matrix $\tilde{\chi}(t)$ for the system and the reservoir is governed by the equation of motion
	\begin{equation}
		\partial_t\tilde{\chi} = -i[\tilde{H}_I(t),\tilde{\chi}], \label{eq:master_full}
	\end{equation}
	where $\tilde{H}_I(t)=e^{i(H_S+H_B)t}H_Ie^{-i(H_S+H_B)t}$ denotes the system-reservoir coupling in the interaction picture, given by
	\begin{equation}
		\tilde{H}_I(t) = \sum_{n} [B_n^\dagger(t) \tilde{c}_n(t) \tilde{c}_{n+1}(t)   + B_n(t) \tilde{c}_n^\dagger(t) \tilde{c}_{n+1}^\dagger(t)].
	\end{equation}
	Here, $B_n(t) = \sum_{k}g_{k,n}^*b_ke^{-i\delta_kt}$ with $\delta_k=\epsilon_k-2V_3$ denotes the Langevin noise operator, and $\tilde{c}_n(t)=e^{iH_0t}c_ne^{-iH_0t}$ with $H_0=\sum_n[J(c_{n+1}^\dagger c_{n}+\mathrm{H.c.}) - V_3 c_{n+1}^\dagger c_{n}^\dagger c_{n} c_{n+1}]$. It is important to note that we consider the collision where the quasiparticles are initially separated from each other, such that both $B_n(t)$ and $\tilde{c}_n(t)$ are slowly-varying operators, whose time scales of the evolution are mainly determined by the hopping strength $J$. Then, by formally integrating Eq.~\eqref{eq:master_full} and substituting it back, we obtain
	\begin{equation}
		\partial_t\tilde{\chi} = -\int_0^t d\tau[\tilde{H}_I(t),[\tilde{H}_I(\tau),\tilde{\chi}(\tau)]]. 
	\end{equation}
	We now trace out the environment to derive an equation of motion for the scar-space density matrix $\tilde{\rho}(t)=\mathrm{Tr}_B[\tilde{\chi}(t)]$. Applying the Born approximation $\tilde{\chi}(t)=\tilde{\rho}(t)\otimes\tilde{\rho}_B$ with $\tilde{\rho}_B=\ketbra{0}_B$, we obtain
	\begin{equation}
	\partial_t\tilde{\rho} = -\sum_{m,n}\int_0^t d\tau G_{mn}(t-\tau)[\tilde{c}_{m}^\dagger(t)\tilde{c}_{m+1}^\dagger(t),\tilde{c}_{n}(\tau)\tilde{c}_{n+1}(\tau)\tilde{\rho}(\tau)]+\mathrm{H.c.},\label{eq:master2}
	\end{equation}
	where $G_{mn}(t)=\langle B_m(t) B_n^\dagger(0)\rangle=\sum_k g_{k,m}^*g_{k,n} e^{-i\delta_kt}$ is the reservoir correlation function. Numerically, we find that the off-diagonal correlator ($m\neq n$) is negligible compared with the diagonal ones ($m=n$), due to the chaotic feature of the thermal-space dynamics. 
\begin{figure}
	\centering
	\includegraphics[width=\linewidth]{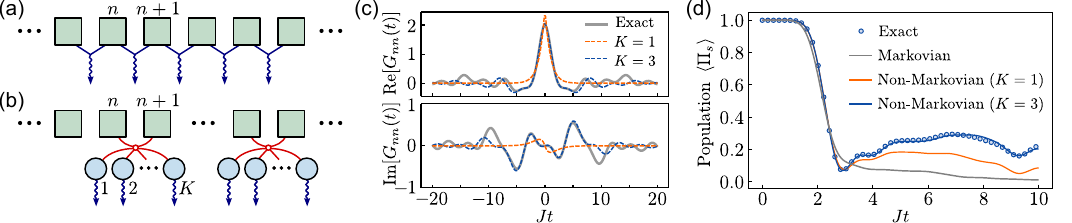}% Here is how to import EPS art
	\caption{(a) Illustration of the Markovian model. (b) Illustration of the non-Markovian model. (c) Real and imaginary part of the correlation function $G_{nn}(t)$ for collisions at the center $n=(L-1)/6$ with $L=19$. The dashed curves show the fitting results with $K=1$ (orange) and $K=3$ (blue) auxiliary modes. (d) Evolution of the scar-space population $\langle\Pi_s\rangle$ for $V_3=0$.} \label{fig:fig_s6}
\end{figure}	
	
	To further simplify the description, we apply the Markovian approximation, where the damping of the correlation function $G_{nn}(t)$ is assumed to be much faster than the evolution of the slowly-varying term $\tilde{c}_n(\tau)\tilde{c}_{n+1}(\tau)\tilde{\rho}(\tau)$ that varies on the time scale $\sim 1/J$. For consistency, we thus take $\tilde{c}_n(\tau)\tilde{c}_{n+1}(\tau)\tilde{\rho}(\tau)$ out of the integral in Eq.~\eqref{eq:master2} and evaluate the integral in a duration $1/J$, i.e.,
	\begin{equation}
	\int_0^t d\tau G_{nn}(t-\tau) \approx \int_{t-1/J}^t d\tau G_{nn}(t-\tau)=\frac{\Gamma_n}{2} + i\delta V_n,
	\end{equation}
	where $\Gamma_n$ and $\delta V_n$ are real numbers and denote the damping rate and the Lamb shift, respectively. Transforming the dynamics back to the Schr{\"o}dinger picture, the reduced density matrix $\rho(t)$ is governed by a master equation in the Lindblad form:
	\begin{equation}
		\partial_t\rho=-i\left[H_S + \sum_n\delta V_nc_{n+1}^\dagger c_n^\dagger c_n c_{n+1},\rho\right]+\sum_n\Gamma_n\left(c_n c_{n+1}\rho c_n^\dagger c_{n+1}^\dagger- \frac{1}{2}\left\{c_n^\dagger c_{n+1}^\dagger c_n c_{n+1},\rho\right\}\right).\label{eq:master}
	\end{equation}
In such a Markovian model, the leakage from the scar space is mediated by the two-particle loss term $\propto \sqrt{\Gamma_n}c_nc_{n+1}$, as illustrated in Fig.~\ref{fig:fig_s6}(a). A typical correlation function $G_{nn}(t)$ for $L=19$ is shown in Fig.~\ref{fig:fig_s6}(c), which indeed exhibits features of a delta function corresponding to a Markovian white noise. As a result, damping of the scar-space probability $\langle \Pi_s\rangle$ can be qualitatively captured by the Eq.~\eqref{eq:master} [see Fig.~\ref{fig:fig_s6}(d)], where $\Pi_s = \sum_{m,n} \ketbra{\phi_{m,n}}{\phi_{m,n}}$ denotes the projector onto the scar space spanned by all the two-quasiparticle states $\ket{\phi_{m,n}}=c_m^\dagger c_n^\dagger \ket{0}$.

We next develop a pseudo-mode approach \cite{garraway1997nonperturbative} that can effectively mimic the non-Markovian effect. As sketched in Fig.~\ref{fig:fig_s6}(b), we assume that each two-body loss $c_n c_{n+1}$ is mediated by a few auxiliary bosonic modes $a_{\sigma,n}$ ($\sigma=1,2,\cdots,K$) decaying to independent Markovian environments. Evolution of the density matrix $\rho_{SA}$ for the system and the auxiliary modes is then governed by the master equation
\begin{align}
	\partial_t\rho_{SA}=&-i\left[H_S+H_A+H_{SA}, \rho_{SA}\right]+\sum_n\kappa_{\sigma,n}\left(a_{\sigma,n}\rho a_{\sigma,n}^\dagger- \frac{1}{2}\left\{a_{\sigma,n}^\dagger a_{\sigma,n},\rho\right\}\right), \hspace{60pt} \\
\textrm{where}\hspace{80pt}	&H_{A}=\sum_{\sigma,n} \Delta_{\sigma,n} a_{\sigma,n}^\dagger a_{\sigma,n},\qquad H_{SA}=\sum_{\sigma,n} (\lambda_{\sigma,n} a_{\sigma,n}^\dagger c_n c_{n+1}  + \mathrm{H.c.}).	\hspace{60pt}
\end{align}
The reduced density matrix of the system can be obtained by tracing out the auxiliary modes, i.e., $\rho(t) = \mathrm{Tr}_A[\rho_{SA}(t)]$. In such a non-Markovian model, the parameters of the auxiliary modes include the detunings $\Delta_{\sigma,n}$, the coupling strengths $\lambda_{\sigma,n}$, and the decay rates $\kappa_{\sigma,n}$. To determine these parameters, we note that these pseudomodes coupled to the Markovian environments can be effectively treated as engineered reservoirs \cite{garraway1997nonperturbative} that feature a correlation function $G_{nn}(t)$ in the form
\begin{equation}
	G_{nn}(t) = \sum_{\sigma=1}^{K} \lambda_{\sigma,n}^2 e^{-\kappa_{\sigma,n}|t|-i\Delta_{\sigma,n}t}.\label{eq:corr}
\end{equation}
The parameters $\{\Delta_{\sigma,n},\lambda_{\sigma,n},\kappa_{\sigma,n}\}$ can thus be extracted by fitting the correlation function $G_{nn}(t)$ according to Eq.~\eqref{eq:corr}. As shown in Fig.~\ref{fig:fig_s6}(d), a few auxiliary modes ($K=3$) suffice to mimic the reservoir correlation function and yield quantitative predictions for the system dynamics [see Fig.~\ref{fig:fig_s6}(d)].

	\section{A\lowercase{pproximate scars from the inelastic quasiparticle collision}}\label{sec:sec5}
	The two-quasiparticle states form a subspace of dimension $\mathcal{O}(L^2)$, which  only has $\mathcal{O}(L)$ couplings to the thermal space. As a result, the subspace spanned by the two-quasiparticle states feature small leakages and can be regarded as a scar space supporting approximate quantum many-body scars (QMBS).
	
\begin{figure}
		\centering
		\includegraphics[width=\linewidth]{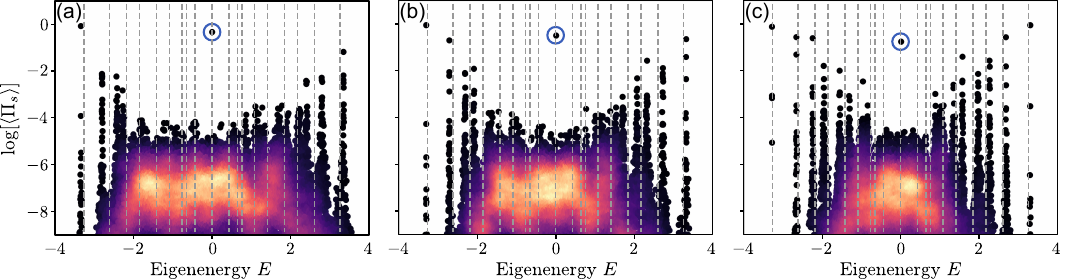}% Here is how to import EPS art
		\caption{Logarithmic plot of the scar-space population, $\log[\langle \Pi_s \rangle]$, evaluated for each eigenstate with energy $E$, for (a) $\xi=1$, (b) $\xi=0.8$, and (c) $\xi=0.6$ for $L=19$. The dashed lines indicate the eigenenergies $\epsilon_k$ of the effective Hamiltonian $H_S$.} \label{fig:fig_s7}
	\end{figure}

	To justify such a physical interpretation, we diagonalize the Hamiltonian for the Hilbert space generated from the inelastic quasiparticle collision, and compute the overlap of each eigenstate $\ket{E}$ (at eigenenergy $E$) with the scar space $\bra{E}\Pi_s\ket{E}$. As shown in Fig.~\ref{fig:fig_s7}(a), several eigenstates have atypical, large overlaps with the scar space, showing clear deviations from the main spectrum and forming visible spikes. These are general features of approximate QMBS, e.g., scarred states in the Rydberg PXP model that have large overlaps with some product states \cite{turner2018quantum}. The locations of these spikes are also close to the eigenenergies $\epsilon_k$ obtained by diagonalizing the scar-space Hamiltonian $H_S$ (dashed gray lines). Visibility of these approximate QMBS can be tuned by  considering a modified Hamiltonian $H_\mathrm{eff}(\xi) = \xi H^{[1]}+H^{[2]}$ with a parameter $\xi\leq 1$ that controls the coupling strength between the scar space and the thermal space. As $\xi$ decreases, the approximate scars show larger overlaps with the scar space, and the spikes become more pronounced and gradually approach the eigenenergies $\epsilon_k$. In addition to approximate QMBS, we also identify one zero-energy eigenstate induced by many-body caging \cite{tan2025interference,ben2025many,nicolau2025fragmentation,jonay2025localized}, which is indicated by the blue circle in Fig.~\ref{fig:fig_s7} and takes the (unnormalized) form
    \begin{equation}
     \ket{\mathrm{zero}}=\xi\sum_{n=0}^{N-2}(-1)^nc_n^\dagger c_{n+2}^\dagger\ket{0} - \sigma_2^-\sigma_3^+(J_-)^{N-1}\ket{\psi_2} -
     \sigma_{L-2}^-\sigma_{L-3}^+(J_+)^{N-1}\ket{\psi_2}.
    \end{equation}

\bibliography{Reference}